# Perspectives and challenges in bolide infrasound processing and interpretation: A focused review with case studies

**Elizabeth A. Silber[1]**

[1]Geophysics, Sandia National Laboratories, Albuquerque, NM, 87123; esilber [at] sandia.gov

**Abstract:** Infrasound sensing plays a critical role in the detection and analysis of bolides, offering passive, cost-effective global monitoring capabilities. Key objectives include determining the timing, location, and yield of these events. Achieving these goals requires a robust approach to detect, analyze, and interpret rapidly moving elevated sources such as bolides (also re-entry and space debris). In light of advancements in infrasonic methodologies, there is a need for a comprehensive overview of the characteristics that distinguish bolides from other infrasound sources, and methodologies for bolide infrasound analysis. This paper provides a focused review of key considerations and presents a unified framework to enhance infrasound processing approaches specifically tailored for bolides. Three representative case studies are presented to demonstrate the practical application of infrasound processing methodologies and deriving source parameters, while exploring challenges associated with bolide-generated infrasound. These case studies underscore the effectiveness of infrasound in determining source parameters and highlight interpretative challenges, such as variations in signal period measurements across different studies. Future research should place emphasis on improving geolocation and yield accuracy. This can be achieved through rigorous and systematic analyses of large, statistically significant samples of such events. aiming to resolve interpretative inconsistencies and explore the causes for variability in signal periods and back azimuths. The topic described here is also relevant to space exploration involving planetary bodies with atmospheres, such as Venus, Mars, and Titan.

**Keywords:** bolide, meteoroid, meteor, asteroid, infrasound, space exploration, signal analysis, planetary defense, re-entry, space debris

## 1. Introduction

Infrasound sensing is a passive and inexpensive global monitoring tool used in the detection and characterization of a wide range of impulsive sources, including bolides. As infrasonic methodologies undergo continuous refinement and evolution toward greater utility and application in planetary defense efforts, there is a pressing need for an up-to-date overview of the factors that distinguish bolides from other sources of infrasound, in addition to a contemporary synthesis of the stages and methodology involved in processing bolide infrasound signals (detection, localization, analysis, interpretation). This paper aims to address these key points by providing an overview of crucial considerations and presenting a unified framework to advance infrasound processing approaches tailored specifically for the study of bolides. Although this paper is geared toward bolides, the content presented here can apply to re-entry of spacecraft and orbital debris. It should be noted, however, that re-entry and orbital debris are significantly slower than bolides and have slightly different considerations in terms of shock physics (Silber et al., 2018).

Meteoroids enter the Earth's atmosphere at speeds from 11.2 – 72.8 km/s (Ceplecha et al., 1998), corresponding to Mach numbers ($M$) between 35 and 270. Mach number is the ratio of the object's speed ($v$) and the local speed of sound ($c(z)$) at a given altitude ($z$), $v/c(z)$ (Ben-Dor et al., 2000; Silber et al., 2018). Upon entering the atmosphere, these objects are subjected to high frictional forces and extreme temperatures (kinetic temperatures readily exceeding 10000 K), resulting in sputtering, ablation, and fragmentation, as well as dissociation and ionization. The resulting luminous phenomenon is known as a meteor. According to the International Astronomical Union (IAU), meteoroids range in size from submillimeter dust particles to less than one meter, while asteroids are defined as objects one meter or larger. Particularly bright meteors, exceeding the absolute visual magnitude (distance of 100 km) of Venus (-4), are classified as fireballs and bolides. The latter often lead to explosive phenomena. Objects brighter than absolute visual magnitude -17 are referred to as superbolides.

Sufficiently large and fast meteoroids generate shock waves that eventually decay to low frequency (<20 Hz) sound waves or infrasound (Silber et al., 2018; Tsikulin, 1970), and depending on the source parameters (e.g., size, velocity) as well as atmospheric conditions along the propagation path, could be detected by microbarometers at large distances (ReVelle, 1976). The meteor-generated shock wave is generally amplified by strong ablation, which is one of the main differences between the natural hypersonic and artificial re-entry objects. For a shock wave to form, the meteoroid must enter the continuum flow regime, characterized by a Knudsen number ($Kn$) of <0.01. Knudsen number ($Kn = \lambda/L$) is the ratio of the mean free path ($\lambda$) of air molecules, and the characteristic length scale ($L$) of the object. However, empirical evidence suggests that under conditions of intense ablation rates and high Mach numbers, meteoroids can produce shock waves at altitudes significantly higher than expected using classical methodologies (Moreno-Ibáñez et al., 2018). This phenomenon, known as hydrodynamic shielding (Bronshten, 1983; Popova et al., 2001; Popova et al., 2000), allows strongly ablating meteoroids to locally reduce their Knudsen number, placing them in the continuum flow regime at higher altitudes (Moreno-Ibáñez et al., 2018; Silber et al., 2018). As a result, it is occasionally possible to detect shock waves originating from very high altitudes (>100 km) (Brown et al., 2007; Silber and Brown, 2014).

Although the topic of seismics is beyond the scope of this paper, it is worthwhile to mention that in certain scenarios, infrasound carries enough energy to induce seismic waves known as air-coupled or atmospheric seismic waves (Edwards et al., 2008). Bolide-generated seismic signals are generally weaker and shorter-lived compared to those associated with earthquakes or other significant seismic phenomena. For further details on this topic, the reader is referred to Edwards et al. (2008).

Energetic bolides can pose a significant threat to life and infrastructure (Trigo-Rodríguez, 2022). In 1908, an airburst over Tunguska, Siberia marked the first documented bolide-generated infrasound, which was detected as far away as England by microbarometers that were invented only five years earlier. As per witness reports, the resulting blast wave was physically felt many tens of kilometers away (Whipple, 1930). While the ground damage was significant, it occurred over a remote, uninhabited area (Chyba et al., 1993; Whipple, 1930). As recently as a decade ago, a large bolide caused injuries to people and significant damage to infrastructure as it traversed the sky over Chelyabinsk, Russia at a shallow angle. The 18 meters diameter asteroid deposited energy of nearly half a megaton of TNT equivalent (1 kt = 4.184E12 J) (Brown et al., 2013; Le Pichon et al., 2013; Pilger et al., 2015; Popova et al., 2013). Coming from the direction of the Sun, the asteroid could not be detected *a priori*, underscoring the unpredictable nature of potentially hazardous objects. More alarmingly, if a bolide is above some critical size, composition and velocity threshold where it poses a physical threat on the ground, there is very little that can be done currently to mitigate the consequences of the ground impact or lower atmospheric airburst (e.g., the Chelyabinsk event) (e.g., Bender et al., 1995; Trigo-Rodríguez, 2022).

Considering that bolides are rapidly moving sources traversing different layers of the Earth's atmosphere in a matter of seconds, they often present an observational challenge in terms of collecting reliable and well-constrained ground truth that would complement infrasound. Moreover, initial parameters of bolides are generally poorly constrained, adding further complexity. This in turn requires a more subtle approach when working with low frequency acoustic signatures from these events.

This paper aims to consolidate methodologies and approaches for bolide infrasound processing and interpretation and present some of the associated challenges. Three representative case studies will be used to demonstrate the practical application of these methods, providing guidelines for consistent use. This topic also has a significant relevance to space exploration, especially in the context of planetary bodies with atmospheres, such as Mars (e.g., Daubar et al., 2023; Fernando et al., 2022; Garcia et al., 2022), Venus (e.g., Blaske et al., 2023; Krishnamoorthy and Bowman, 2023), and Titan (e.g., Bowman, 2021; Petculescu and Lueptow, 2007).

This focused review paper is organized as follows: Section 2 provides a brief overview of various considerations related to bolides and bolide-generated infrasound; Section 3 focuses on global detections of bolides and the case studies; in Section 4, methodology to detect and process infrasound are given; the analyses of the case studies and discussion are presented in Section 5, and the conclusions and future work in Section 6.

## 2. Key Considerations in Bolide Infrasound

### 2.1 Modes of Shock Production

Bolides generate shock waves, and consequently infrasound, through two primary mechanisms: a hypersonic passage through the atmosphere, and fragmentation (Figure 1). Due to their hypervelocity, meteoroids create a narrow Mach cone (<2°), which can be approximated as a cylindrical line source with a ballistic shock wave expanding radially from the trajectory (ReVelle, 1976) (Figure 2). However, depending on various parameters (e.g., ablation rate, altitude), highly non-linear effects due to intense heating and ablation take place, which can distort the shock front such that it propagates outward at angles up to 25° from normal (Brown et al., 2007; Silber and Brown, 2014; Zinn et al., 2004). Fragmentation can occur as discrete fragmentation episodes or continuous fragmentation (Trigo-Rodríguez et al., 2021) (Figure 1). Depending on the physical properties of an object, some events might have no fragmentation while others might undergo one or both types of fragmentation. Terminal fragmentation that results in disintegration of the object is known as an airburst (Figure 1). In theory, the shock wave originating from a fragmentation event should have a different morphology from a typical cylindrical shock wave. That shock is considered a quasi-spherical or quasi-point source, because there is generally some component of forward momentum, which differentiates these kinds of events from a fully stationary point-source such as a high-altitude nuclear explosion.

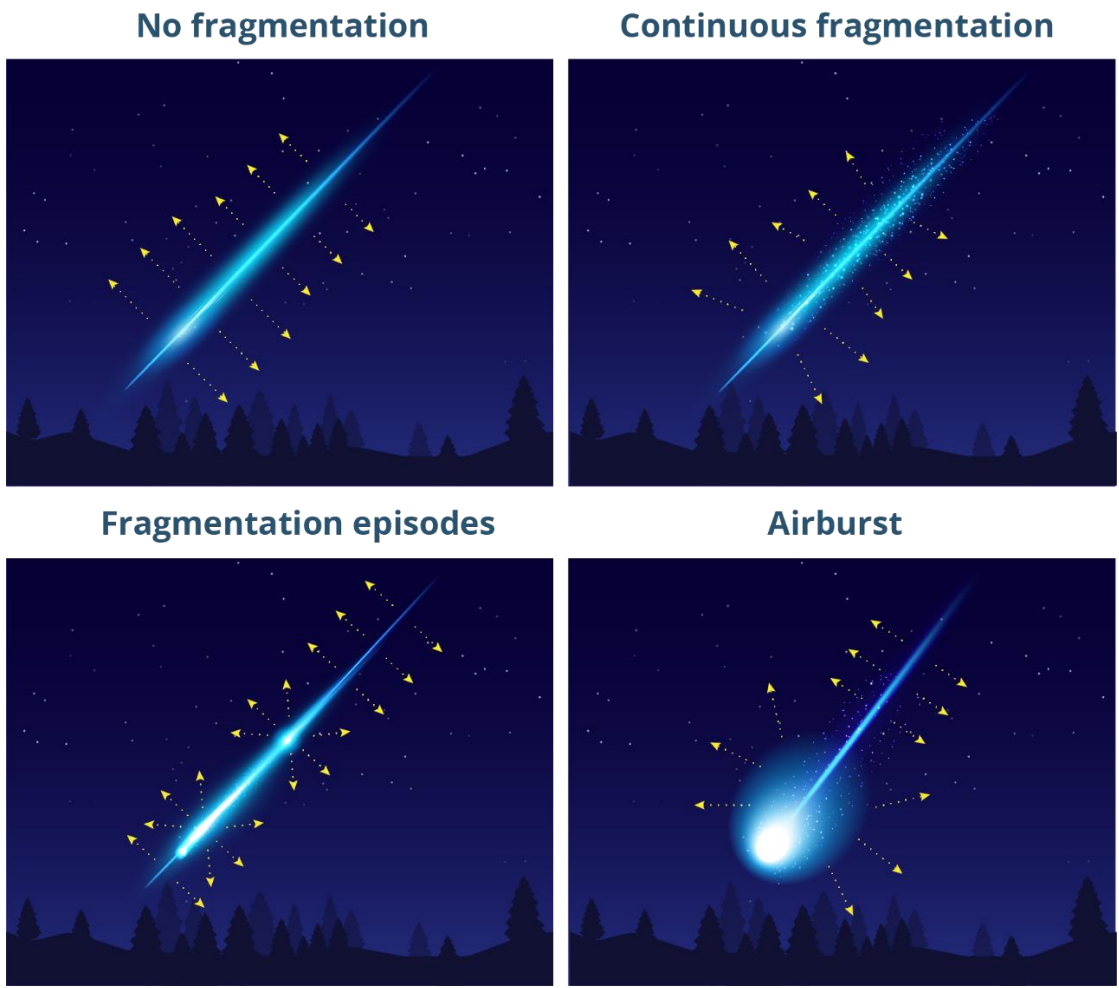

**Figure 1.** Diagram showing various modes of shock production. The arrows broadly indicate the direction of shock propagation, although the real-life events are generally more complex.

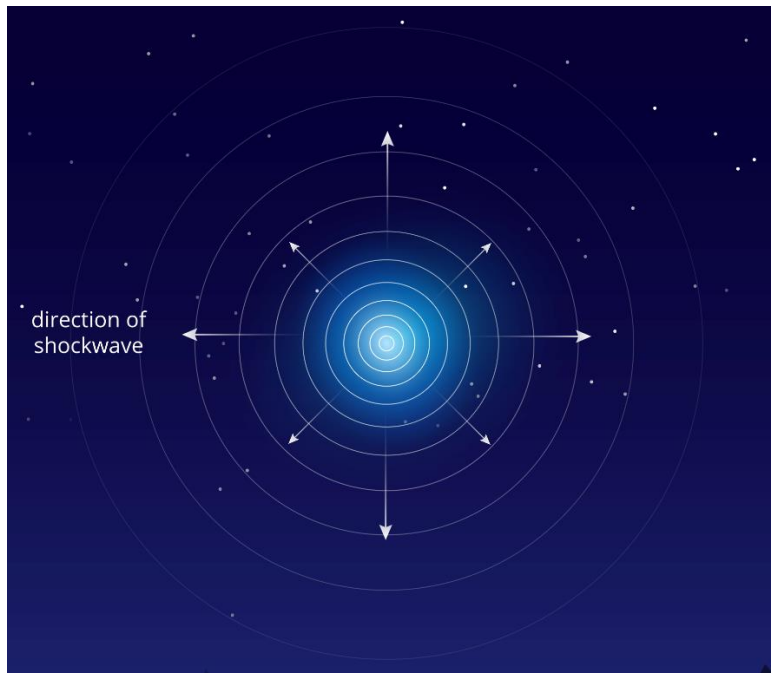

**Figure 2.** Diagram showing a simplified head-on view (in the direction of propagation) of a bolide. The arrows depict the radial direction of shock propagation. The bright region indicates where the strong shock occurs. The shock pattern in a real-life scenario can be more complex.

### *2.2 Considerations in Short- and Long-Range Detections*

Unlike stationary sources, bolides release energy over their path length, necessitating careful interpretation. The type of shock and the trajectory's geometry in relation to the stations are fundamental factors affecting signal detectability. Typical flight trajectories of bolides may extend for tens or even hundreds of kilometers depending on factors like an entry angle (e.g., Moreno et al., 2016; Shober et al., 2020). The trajectory length ($d_{train}$) has a significant impact on the detection and localization of these events using infrasound in near-field. Predicted back azimuths (direction of the signal arrival at the station) and signal travel times can have large variations depending on the bolide trajectory orientation with respect to the station (see Figure 3). Moreover, the nature of the shock (ballistic vs. quasi-spherical) also governs the detectability of the signal in both near- and far-field. Figure 3 shows an idealized diagram of the bolide shock geometry relative to two infrasound stations. It presents two scenarios: no fragmentation (left panel) and fragmentation (right panel), with the latter illustrating one of several possible variations. In the case of ballistic shock geometry, Station A, positioned to the side and below the trail, can capture signals from a specific point along the trajectory. Station B is less likely to detect a cylindrical line source but could register a fragmentation event if it occurs with a favorable geometry. Station A could hypothetically receive two signals, one from a cylindrical line source, and one from gross fragmentation. An illustrative practical example of how infrasound signal detections depend on the source geometry is the recent re-entry of OSIRIS-REx (Origins, Spectral Interpretation, Resource Identification, and Security – Regolith Explorer) sample return capsule (Fernando et al., 2024a; Fernando et al., 2024b; Silber et al., 2024).

Interpreting signals can be further complicated by other factors. For instance, signals from higher altitudes might be detected later than those from lower altitudes (Brown et al., 2011). Due to shock distortion, ballistic arrivals may deviate from the shock normal by up to 25° (Brown et al., 2007). Fragmentation events at different altitudes may each produce a shock detected by the same station (Brown et al., 2011; Silber and Brown, 2014). Additionally, signal multipathing can occur due to atmospheric propagation effects (e.g., Hedlin and Walker, 2013). Sometimes signal multipathing as well as signals generated by two or more points along the trail can be present in the same waveform, complicating analysis and interpretation. Accurate ground truth information, such as trajectory and timing, is therefore essential for effective interpretation.

On the other hand, distant events, where the trajectory length ($d_{traj}$) is negligible relative to the total distance ($d$) between the event and an infrasound station ($d_{traj} << d$), can be treated as point-sources (ReVelle, 1976). However, this is an oversimplification, as it is still possible to infer the shock altitude, albeit with great difficulty (e.g., Silber et al., 2011). Moreover, even at long ranges, there may still be discernible directional patterns in the emission of acoustic energy (Pilger et al., 2019; Pilger et al., 2015). For a comprehensive overview of bolide infrasound, the reader is directed to a book chapter by Silber and Brown (2019).

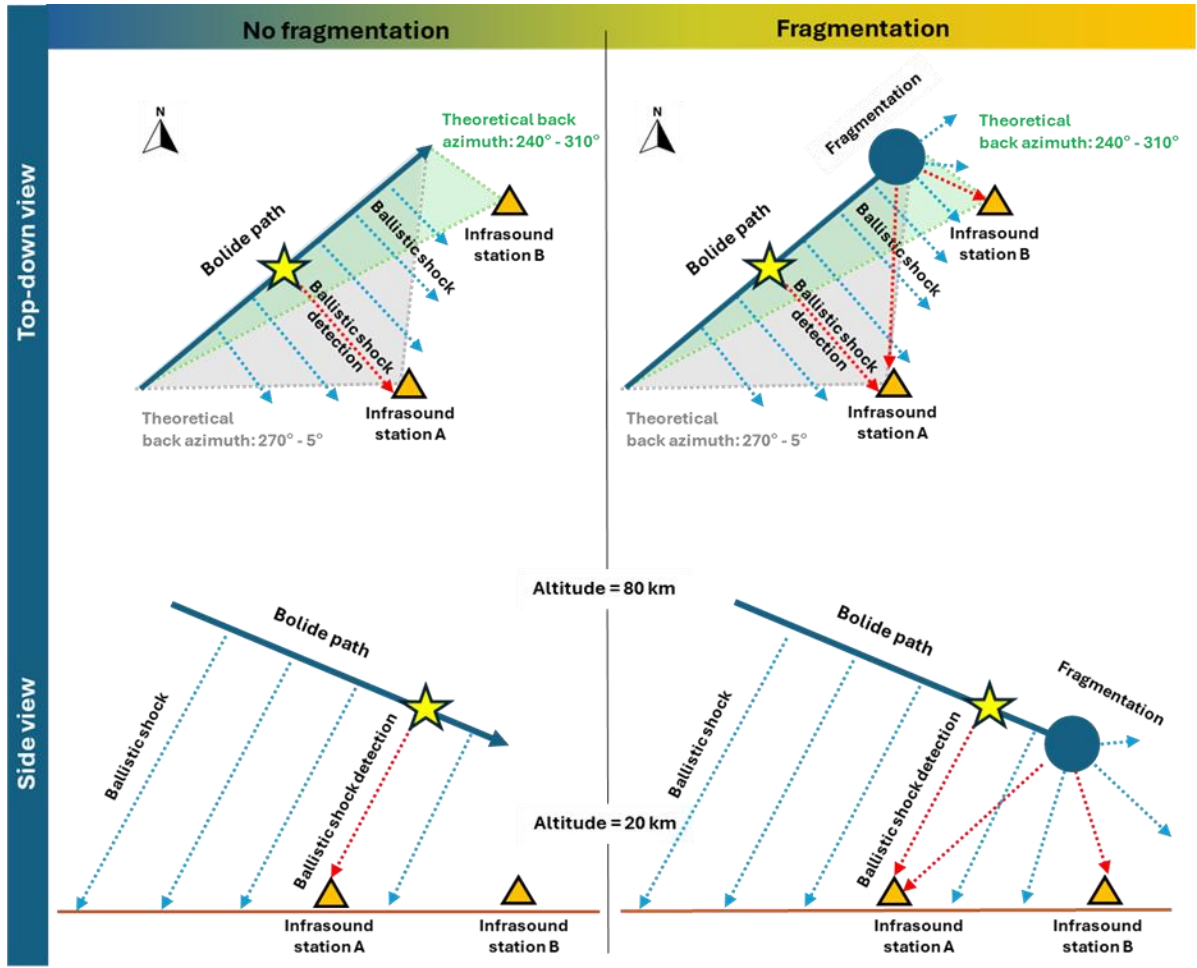

**Figure 3.** Diagram showing short-range infrasound propagation considerations for detecting bolides. Top down (top panels) and side view (bottom panels) show an idealized representation of a bolide trajectory (arrows) relative to two infrasound stations (triangles), and different types of shock (ballistic versus spherical). Depending on the orientation and the mode of shock production, any one station might detect signals from a different part of the trail (the stars represent the ballistic signal origin). In the example given for fragmentation (circles), the station might receive two signals.

## 2.3 Energy Deposition

Through their hypersonic flight, bolides deposit energy as a function of path length, where the highly non-linear region is known as a blast radius or characteristic radius ($R_0$). The definition of $R_0$ for cylindrical line sources (no fragmentation) is:

$$R_0 = \left[\frac{E_l(z)}{p_s(z)}\right]^{1/2} \approx M d_i \tag{1}$$

where ($E_L$) is the energy deposited per unit length, $p_s(z)$ is the ambient pressure at source altitude ($z$), and $d_i$ is the impactor diameter.

A spherical shock formed by fragmentation will have its the characteristic radius proportional to the cube root of the energy ($E_S$) released by the explosion:

$$R_0 = \left[\frac{E_S(z)}{\frac{4\pi}{3}p_s(z)}\right]^{1/3} \qquad (2).$$

The highly non-linear shock will transition to a weak shock beyond approximately $10R_0$, with the fundamental frequency ($f_0$) defined as:

$$f_0 = \frac{1}{\tau_0} = \frac{c_s(z)}{2.81R_0} \qquad (3),$$

where $\tau_0$ is the fundamental period of the wave (ReVelle, 1976). The wave will eventually transition into a linear regime as attenuation and dispersion act to reduce the amplitude and 'stretch' the period (Dumond et al., 1946). Some fraction of energy deposited by a bolide will ultimately go into acoustic energy of the remnant shock, typically at infrasonic frequencies. Once in the linear regime, the dominant signal period is assumed to be constant. In theory, the dominant signal period can be used to infer energy deposition by a bolide at a particular point along the trajectory. It has been found that dominant signal periods can vary significantly from station to station, which is generally attributed to signal being generated at different points along the trail and subsequently 'sampled' by different stations. Other factors, albeit with a lesser effect, are station noise and range dependence (observed signal periods increase as a function of range due to attenuation, with longer-period signals experiencing lower attenuation compared to shorter-period signals).

Empirical energy relationships are widely used for estimating yield, yet many were originally developed for different types of explosions, such as nuclear and chemical blasts. For example, in the mid-20[th] century, the Air Force Technical Applications Center (AFTAC) derived period-yield relations for nuclear explosions (Revelle, 1997). The relations use the dominant signal period as the primary variable, which has been found to be relatively robust for stationary, point-source aerial explosions due to minimal expected variation in dominant signal period across different observation points. In the late 1990s, Revelle (1997) adapted these relations to bolides by excluding the radiation component. Other energy relationships incorporate signal amplitude and range, although signal amplitude is challenging due to its susceptibility to propagation effects and has been identified as the least reliable parameter for yield estimation. Over time, some relationships have been refined and tailored for bolides. In calculating yield, only stratospheric arrivals should be used. Thermospheric arrivals are unreliable due to heavy attenuation and have not been systematically tested in deriving empirical energy relations. A comprehensive list of energy relations is given in Silber and Brown (2019).

### 2.4 Infrasound Sensing

Infrasound emerged as a critical sensing modality in the mid-20[th] century primarily for monitoring nuclear explosions, a role that diminished following the cessation of atmospheric testing. In recent decades, infrasound sensing has been established as a means for passive and inexpensive global monitoring, and it currently serves toward that goal as one of the four technologies of the International Monitoring System (IMS) of the Preparatory Commission of the Comprehensive Nuclear-Test-Ban Treaty Organization (Brachet et al., 2010; Christie and Campus, 2010). The IMS comprises a global network encompassing 53 certified stations (out of a planned 60), strategically positioned to detect illicit explosions on a global scale (see Figure 4). The IMS is optimized to identify and detect a 1kt explosion anywhere on the globe (National Research Council, 2012). Infrasound monitoring, characterized by its passive nature, extends beyond its primary role to encompass a wide array of civilian uses, including the detection of bolides (Pilger et al., 2020).

The continuous, global monitoring capabilities of infrasound provide a layer of detection that can complement or even augment other sensing modalities and provide independent observations in absence of well-constrained ground truth. Unlike optical observations, infrasound sensing is less sensitive to diurnal cycles and is unaffected by cloud coverage conditions, which makes it an indispensable tool for detection of bolides (e.g., Arrowsmith et al., 2008; Ott et al., 2019; Pilger et al., 2019; Pilger et al., 2020; Silber et al., 2011). It should be noted, however, that background noise levels are generally higher during the day due to increased winds, which could adversely affect the detectability (also see Section 2.5).

The design of infrasound sensors and station configuration has been extensively documented in the literature (Brachet et al., 2010; Christie and Campus, 2010; Mutschlecner and Whitaker, 1997; Slad and Merchant, 2021), and therefore, only their most relevant application to bolides will be addressed here. Similar to other sources of infrasound, signals from bolides and re-entry events can be detected using both arrays and single sensor stations. Arrays, which consist of three or more sensors arranged in various configurations, are particularly advantageous because they allow for determining the direction of signal arrival (back azimuth) and how steep the airwave sweeping the array is (trace velocity). Besides ground-based stations, in recent years there has been an emergence of high-altitude balloon-borne platforms (Albert et al., 2023; Bowman and Albert, 2018; Bowman et al., 2022). Sensors suspended on high-altitude balloons floating in the stratosphere provide a unique vantage point as they can capture signals that are efficiently channeled through the stratospheric propagation duct (AtmoSOFAR) (Albert et al., 2023). More recently, balloons, in addition to ground-based stations, have been utilized to observe the re-entry of the NASA's (National Aeronautics and Space Administration's) OSIRIS-REx sample return capsule (Silber et al., 2024).

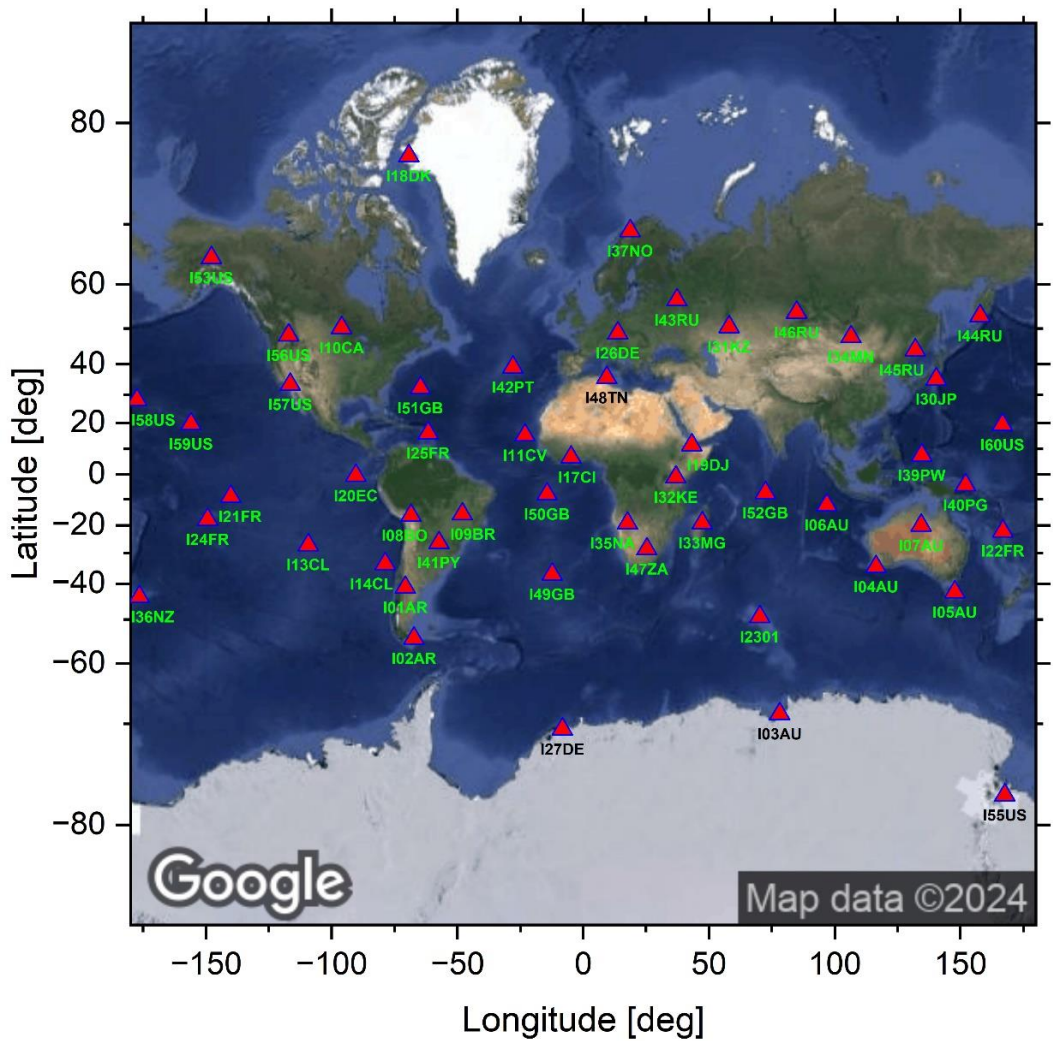

**Figure 4.** Certified infrasound stations of the IMS. Map source: Google.

### *2.5 Propagation Effects*

An important aspect of infrasound is the effect of the atmosphere on signal propagation (e.g., de Groot-Hedlin et al., 2010; Drob, 2019; Drob et al., 2010; Pilger et al., 2019; Pilger et al., 2015). A comprehensive exploration of this topic necessitates a dedicated paper, exceeding the scope of the current study. Therefore, only pertinent highlights will be mentioned herein, with references provided for further exploration by the reader.

The effective speed of sound ($c_{eff}$) in the atmosphere is a function of temperature ($T$) and wind speed along the propagation path, given by: $c=(\gamma RT/M)^{1/2} + n \cdot v_w$, where $\gamma$ is the specific heat ratio, $R$ is the molar gas constant, $M$ is the molar mass of the gas, n is the unit vector and $v_w$ is the wind speed along the propagation path. Due to the refractive nature of acoustic waves, infrasound can be 'ducted' through the atmospheric waveguides which are formed as a result of variations in temperature and wind speed in the atmosphere. At short ranges of up to 300 km, infrasound arrivals can be direct, i.e., the acoustic wave propagates from an elevated source to the receiver in a more-or-less a straight path, without any 'bounces'. These arrivals are sometimes referred to as near-field. Beyond this distance, or far-field, the acoustic wave experiences one or more 'bounces' and can be channeled through different waveguides, resulting in longer travel times, lower amplitudes, and more complex waveforms. There are three primary waveguides: tropospheric (310 – 330 m/s), stratospheric (280 – 310 m/s) and thermospheric (180 – 300 m/s) (Kulichkov, 2000; Negraru et al., 2010; Nippress et al., 2014). Arrivals within the 'boundary layer' (reflection altitudes <1 km) have celerities >330 m/s (Kulichkov, 2000). The celerity is the average speed based on the source-to-receiver distance and time of infrasound signal arrival. The dynamic nature of the atmosphere can adversely affect the infrasound signal as it propagates from source to receiver, even at short distances (e.g., Averbuch et al., 2022a; Drob et al., 2010; Hedlin et al., 2002; Hedlin and Walker, 2013). This variability can lead to unexpected detections in acoustic shadow zones or the absence of anticipated signals (Green et al., 2011; Silber and Bowman, 2023). The influence of prevailing winds can be notable, affecting detection capabilities by enhancing signals downwind and attenuating them upwind, while also potentially altering signal amplitudes and inducing Doppler shifts in signal periods. Additionally, small-scale atmospheric perturbations and scattering effects contribute further variability to the signal characteristics (e.g., Chunchuzov, 2004; Norris et al., 2010; Silber and Brown, 2014). Furthermore, accurately discerning between multipath propagation and shock waves originating from distinct points along the bolide trajectory poses a challenge in signal analysis (Silber and Brown, 2014). Considering that bolides differ from other common sources of infrasound are their speed and the large span of altitudes they traverse, these properties must be carefully considered during analysis and interpretation.

## 3. Global Detections of Bolides

Infrasound can be generated by a variety of sources of anthropogenic and natural origin, including volcanoes (Matoza et al., 2022), explosions (Golden and Negraru, 2011), rocket launches (Pilger et al., 2021), aurora (Wilson, 1971), lightning (Farges and Blanc, 2010), and bolides (Pilger et al., 2019; Silber, 2024). Many natural and anthropogenic phenomena produce infrasound signals with overlapping dominant frequencies and similar dominant signal periods. Thus, without any knowledge of ground truth information, infrasound sensing lacks the robustness needed to accurately identify an event based on waveform records alone. Ground truth generally comes from other sensing modalities, including witness reports, optical, radar, and space assets. There is a growing number of all-sky camera networks around the world; however, their coverage is limited to certain geographical regions.

Observations from space offer global coverage but thus far, there is no system that is specifically dedicated to bolides. The only space assets with global coverage are U.S. Government (USG) sensors, which detect and report bright flashes associated with bolides (Nemtchinov et al., 1997). While very little is known about this system because it is restricted, the metadata reported on the NASA Jet Propulsion Laboratory (JPL) Center for Near-Earth Object Studies (CNEOS) webpage (herein referred to as CNEOS) is indispensable. As of writing this paper, the CNEOS list includes nearly 1000 bolide events dating back to 1988.

Figure 5 shows all bolides detected by USG sensors as a function of energy released. All events include date, time (in UTC), geographical location, and energy released, and most also have an altitude of peak brightness. Over 300 events also include the velocity vector which can be used to tabulate an entry angle and estimate orbital parameters (Peña-Asensio et al., 2022). As of writing this paper, the bolides up to 2022 also include light curves, or brightness as a function of time, which can be used to infer the overall bolide behavior during entry, including fragmentation episodes.

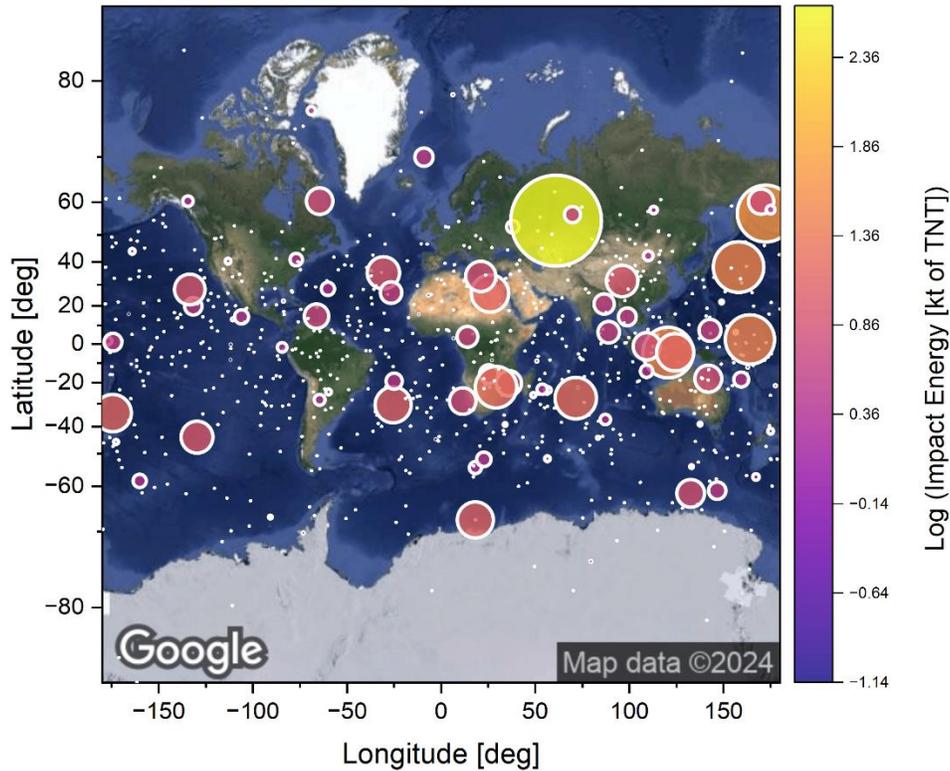

**Figure 5.** All bolide events listed on the CNEOS webpage. Circle size corresponds to the log of impact energy, which is in units of kt of TNT equivalent. For reference, the Chelyabinsk bolide (yellow circle) released energy of 440 kt. Map source: Google.

### 3.1 Case Studies

Three events have been chosen to serve as examples to demonstrate the approach to analyze infrasound and illustrate the associated challenges. Two events are near-field (<300 km) and one is a far-field event.

### 3.1.1 Near-field: A Regional Event and the Greenland Bolide

The first example is a regional event in Southwestern Ontario, Canada, previously published by Silber and Brown (2014). It was selected because of the well-characterized ground truth information, relatively long trajectory (119 km), and its proximity to a local infrasound station. It was detected on January 26, 2009 (07:16:23 UTC) by several regional all-sky cameras which provided excellent constraints on the trajectory, velocity (67.7 km/s), and entry angle (25°). All entry angles given in this paper are relative to the horizontal. This was a high-altitude event, with the luminous path extending from an altitude of 115 km down to 81.7 km. While detections of infrasound from high altitudes are rare, high velocity objects undergoing strong ablation can produce a shock wave at higher altitudes (see e.g., Moreno-Ibáñez et al., 2018; Silber et al., 2018).

The second example is the 2018 bolide over Greenland, which raised attention due to its energetics and the close proximity to I18DK infrasound station. As per the CNEOS report, the bolide entered on July 25, 2018 (21:55:26 UTC, 19:55:26 local time) at a velocity of 24.4 km/s and at a shallow angle of 11°, reaching peak brightness at an altitude of 43.3 km. The direction of the bolide flight was from west to east (Peña-Asensio et al., 2022). The residents of the town Qaanaaq reported brightening of the sky and hearing sonic booms at around 20:00 local time. The CNEOS-reported location is 50 km north from the Pituffik Space Base (previously known as the Thule Air Force Base), which attests to the need to correctly identify natural impacts (e.g., Popova et al., 2013; Silber, 2024). The bolide produced seismic signals (Karakostas et al., 2020; Schmerr et al., 2018) as well as infrasound (Pilger et al., 2020).

### 3.1.2 Far-field: The Indonesian Bolide

The Indonesian bolide (October 8, 2009) remains one of the top three most energetic events listed in the CNEOS database, sharing the third spot with another event of equivalent energy (33 kt). Only the Bering Sea (49 kt) and Chelyabinsk (440 kt) bolides were more energetic. The bolide impacted on October 8, 2009, at 02:57:00 UTC (10:57:00 AM local time) over the province of South Sulawesi, leaving a lingering smoke trail which was captured on video. It took several years before USG sensor data was released; at the time, only infrasound records could be used to estimate energy and establish geolocation (Silber et al., 2011). Through the combination of the observed signal period analysis at stations <5000 km from the source and numerical modeling, Silber et al. (2011) estimated the airburst altitude at ~20 km altitude, an entry velocity of <20 km/s, and an entry angle of 50-80°. According to USG sensor observations published on the CNEOS website an altitude of the airburst was 19.1 km. The bolide entered from north-northwest at a velocity of 19.2 km/s and at an angle of 67°. Peña-Asensio et al. (2022) estimated the diameter at ~5 m and determined the object to be of asteroidal origin (Tisserand parameter = 5.1). This event was selected not only for its energetics but to demonstrate the variability in infrasound signal measurements through different studies and illustrate that event analysis might not always be a straightforward process. Furthermore, it is hypothesized in this work that the bolide entry angle might play a more significant role than previously thought in signal analysis and interpretation. This has implications for detections of artificial object re-entry which typically have shallow entry angles.

## 4. Methodology

### *4.1 Signal Detection and Analysis*

A search for infrasound signals typically starts with some ground truth information. Assuming that the event location and time are known, then it is straight forward to compute a 'detection probability time window' applicable to any fixed station. This calculation hinges on the station's geographical position relative to the source and considers the theoretical propagation speeds of acoustic waves along the shortest path between source and receiver, accounting for all possible waveguide conditions. The fundamental premise is that each potential propagation channel defines a specific time interval during which a systematic search can effectively identify signals from a known event. This practical approach is advantageous when analyzing extensive datasets to detect signals originating from specific point- or moving sources. Essentially, the detection probability time window specifies the segment of waveform duration within which detectable signals may realistically appear, incorporating the earliest feasible arrival times (boundary layer and tropospheric waveguide) and the latest possible arrival times (thermospheric waveguide). Signals detected outside this designated time window cannot reasonably be associated with the event of interest, such as infrasound signals traveling at physically unrealistic speeds (either extremely low or high). Additionally, this computation yields supplementary information such as a theoretical back azimuth, indicating the direction from which the wavefront originates. Nevertheless, it does not inherently establish any association with detected signals, necessitating methods such as beamforming for signal detection.

The next step is to determine if the signal exists within the 'detection probability time window' and if it does, to obtain signal parameters such as its direction of arrival (DOA), which is a vector consisting of the back azimuth and elevation angle (represented through signal trace velocity). There are different techniques used for detection of signals, including beamforming. While a discussion on how these different approaches work is beyond the scope of this paper, it is worth mentioning that there are a variety of software packages available for this purpose, including MatSeis (Harris and Young, 1997), InfraPy (Blom et al., 2016), and DTK-GPMCC (DASE Toolkit – Graphical Progressive Multi-Channel Correlation (PMCC) algorithm) (Cansi, 1995; Mialle et al., 2019). The latter is sensitive to signals with a low signal-to-noise ratio (SNR), which makes it suitable for finding weak signals or dealing with stations with high local noise. Some packages use multiple frequency bands (PMCC) while others (MatSeis) require some initial bandpass filter corner frequencies in order to resolve signals. In MatSeis, unless a bolide is very energetic with signals at very low frequencies, a typical bandpass for an initial search is 0.2 – 3 Hz (Edwards et al., 2006; Ens et al., 2012; Gi and Brown, 2017). However, signals might not always be apparent, requiring tweaking of the beamformer parameters to resolve a possible infrasound signature. Signals might be weak or attenuated due to propagation effects and/or a station might be noisy. Moreover, bolides are not all the same, and despite having similar energies, might exhibit notable intrinsic variations. Therefore, a search for signals can turn into an iterative, trial-and-error process, requiring an analyst to refine parameters at each step before a satisfactory solution is achieved.

The best practice approach to analyze bolide infrasound signals has been established to ensure consistency across different data sets (Edwards et al., 2006; Ens et al., 2012; Revelle, 1997). Edwards et al. (2006) outlined seven steps (plus one optional step for large events) during the pre-processing and analysis stage to ensure consistency and no systematic bias when processing different bolide events. These steps include: (0) apply instrument response correction for large events; (1) find the onset and duration of the signal; (2) calculate the average back azimuth and trace velocity; (3) stack waveforms; (4) derive and measure the maximum signal envelope amplitude; (5) measure the peak-to-peak amplitude and period at maximum amplitude; (6) tabulate the total integrated energy/power and background noise levels; and (7) calculate the integrated SNR. Ens et al. (2012) performed the analysis of 71 bolides detected on 143 IMS stations to derive a set of empirical infrasound source discriminators. In their work, Ens et al. (2012) determined that Step (0) can be omitted as the outcome is negligible, and expanded on Steps (4) and (7) to include a more rigorous and iterative approach to obtain corner cutoff frequencies and perform power spectral density (PSD) analysis, and subsequently extract SNR and integrated signal energy. This will be described later in this section.

The software packages mentioned earlier can extract many different signal parameters (e.g., back azimuth, signal onset and duration, trace velocity, frequency content, number of arrivals/phases, etc.), including quantities listed in Steps (1) and (2). Also, the best beam or 'stacked' waveform, as noted in Step (3), can be extracted for further processing.

Only signals with a back azimuth consistent with that predicted are analyzed further. However, while it is typically assumed that back azimuths should be within approximately ±10° (e.g., Edwards et al., 2006; Ott et al., 2019) for far-field scenarios, in rare occasions deviations of ±25° have been noted (Ott et al., 2019; Silber et al., 2011), and in extreme cases, as much as 30° (Gi and Brown, 2017). Such situations might arise because cross winds along the propagation path might be unusually strong (Diamond, 1964; Green et al., 2018), or signals from a different part of the trajectory might be sampled (Silber and Brown, 2014), especially if the bolide enters at a shallow angle and as a result traverses a long path (e.g., Chelyabinsk). In the near-field, back azimuths might vary drastically depending on the trajectory orientation relative to the station and the modes of shock production (see Figures 1 and 3). Furthermore, caution must be exercised in signal analysis as some variability in apparent back azimuth might arise as a result of analyst's choice of window length, the presence of acoustic coda, and high station noise (or some combination of these). Therefore, unless the ground truth is very well-constrained (i.e., the accurate trajectory is known and fragmentation points identified), one should not readily dismiss any signals without further analysis.

To accomplish Steps (4) – (7), it is necessary to perform a careful analysis of the power spectral density (PSD) of the signal in order to obtain accurate bandpass corner frequencies, establish the dominant frequency from PSD peak, and measure the signal amplitude and period. A detailed description of the PSD analysis can be found in Ens et al. (2012); thus, only the relevant highlights will be given here.

Using the best beam or 'stacked' waveforms from all sensors, the PSD is computed for the entire signal to encompass all frequency components. Subsequently, a smaller time window centered around the peak signal amplitude is analyzed to optimize the SNR. The final window size represents a balance between minimizing the length of the window to avoid low-frequency cutoff issues while maintaining adequate spectral resolution tailored to the signal frequency. Adjustments to this window length should be made based on signal-specific characteristics. For example, events with higher energy, such as energetic bolides with lower frequencies, required longer window durations compared to less energetic events. Additionally, the PSD is computed for both the overall infrasonic waveform and a noise window equivalent in length to the signal's duration, with Hann windowing applied to mitigate cutoff errors in the PSD calculation. Bartlett's method for PSD estimation is preferred over noise-reducing methods like Welch's (Proakis and Manolakis, 2006) to preserve spectral resolution, as it avoids dividing the signal into smaller segments that could compromise frequency precision.

With the aim to accurately determine the signal frequency bandpass, some sacrifice in SNR might be acceptable to enhance spectral clarity. To isolate the signal PSD with minimal noise, the noise PSD is derived as the average of equivalent time windows before and after the signal, subtracted from the total signal's PSD measurement. This approach aims to accurately determine the frequency bandpass containing signal energy (Figure 6), balancing SNR considerations with spectral resolution.

The described methodology represents an iterative process aimed at achieving optimal results and determining the frequency cutoff accurately.

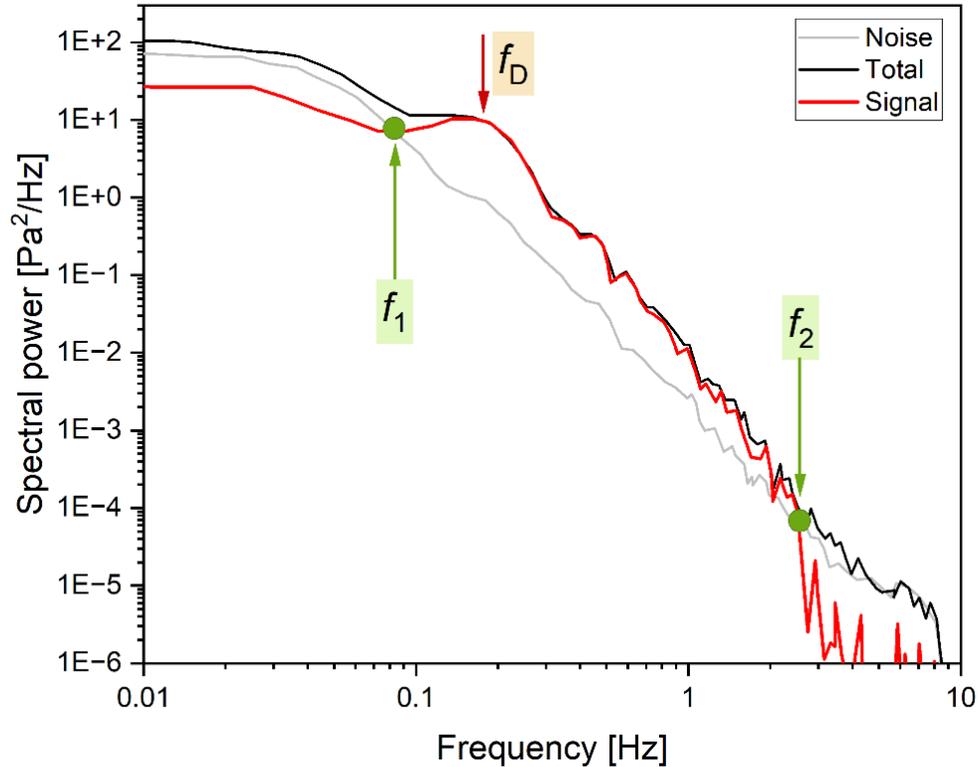

**Figure 6.** A representative example of a PSD showing the signal and where cutoff frequencies are picked.

The amplitude measurements should include peak-to-peak amplitude ($A_{P2P}$) and maximum amplitude ($A_{max}$), both of which are obtained at the same point of the filtered, stacked waveform. First, the amplitude envelope is derived by computing the Hilbert Transform (Dziewonski and Hales, 1972), with the maximum amplitude corresponding to the envelope's peak. The dominant signal period ($T_{ZC}$) is measured at zero-crossings at maximum signal amplitude. The signal period through PSD ($T_{PSD}$) is obtained by inverting the dominant frequency, expressed as $T_{PSD} = 1/f_{max}$.

Ens et al. (2012) added an additional step, which will be defined here as Step (8), and that is to account for a possible Doppler shift in the signal period as infrasound propagates through the stratospheric waveguide. For this purpose, it is useful to obtain the average stratospheric wind velocity between source and receiver over the altitude interval between 40 km and 60 km using range dependent atmospheric profiles along the propagation path. This average wind velocity can also serve as a diagnostic tool for assessing signal strength variations, such as those caused by upwind or downwind propagation. Ens et al. (2012) noted, however, this method of obtaining a single average value may be unreliable due to the standard deviation of the wind velocity often being larger than its mean value, particularly for long-range propagation scenarios.

### *4.2 Signal Characteristics*

Infrasound signals generated by bolides can exhibit a large range of variations, including duration, number of phases (or packets), and other features. Figure 7 shows representative examples of previously analyzed and published infrasound signals detected in near-field and far-field (Silber, 2014, 2024). For example, at short ranges signals will often have an N-wave signature (Silber and Brown, 2014). There could be multiple arrivals (Figure 7d,e) but caution has to be given to determine the cause – atmospheric multipathing (Hedlin and Walker, 2013) or signals coming from different parts of the trail and/or signals generated by different modes of shock production (cylindrical line source and fragmentation) (Brown et al., 2011; Silber and Brown, 2014).

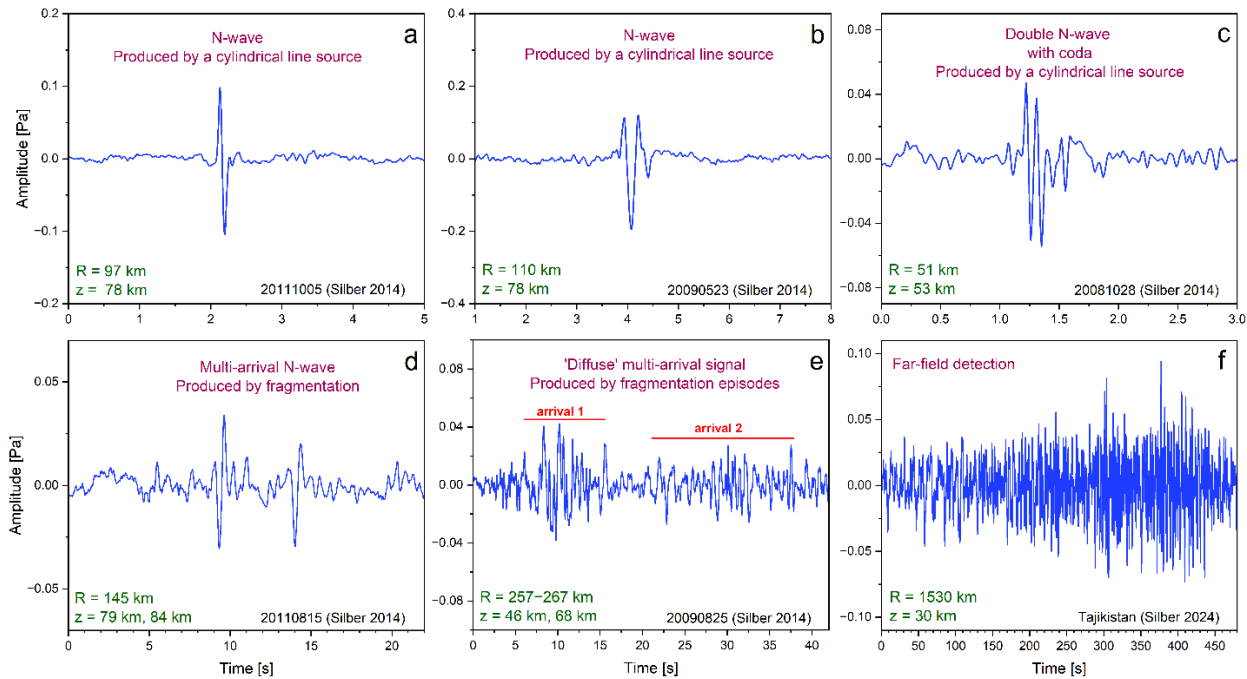

**Figure 7.** Examples of infrasound signals produced by bolides. In each case, the waveform is 'stacked' and filtered. *R* is the horizontal range between source and receiver, and *z* is the source altitude.

*4.3 Propagation Modeling*

Propagation modeling is a powerful approach that can provide additional constraints on observed bolide infrasound signals (or lack thereof). For example, it can be used to investigate the existence of propagation paths and waveguides, compare predicted and observed signal travel times, back azimuth, calculate attenuation in signal amplitude, and constrain the altitude of the shock (e.g., Silber and Brown, 2014; Silber et al., 2011). There are different approaches, from geometric raytracing (e.g., Blom, 2014) to parabolic equation methods (e.g., Waxler et al., 2017). This topic is extensive and warrants a paper of its own; therefore, only relevant highlights will be mentioned here in the context of bolide infrasound.

In the near-field, where the length of the trajectory is non-negligible relative to the distance to a station, propagation modeling can be applied in two ways. If the trajectory is not well constrained and only limited information is known, then in some limited circumstances propagation modeling can be leveraged for reconstructing the shock altitude (e.g., Pilger et al., 2020). However, this is a challenging task and can involve large uncertainties.

The other approach is predicated on the assumption that the trajectory is well-constrained using other observational means (e.g., optical). In such a case, propagation modeling must be done using the entire length of the trail. Since the bolide energy deposition is a function of path length, the trajectory can be assumed to be partitioned into a large number of discrete points (of infinitesimal length), each being a source of infrasound. The infrasonic wave is then propagated from each discrete point to the station, and the results are compared to observations to estimate the likely source of the shock. An example of raytracing outputs for a regional event with a well-constrained trajectory, originally published by Silber (2014), is shown in Figure 8. The observed travel time (Figure 8a) and back azimuth (Figure 8b) are compared to raytracing outputs to estimate the shock source height. Furthermore, ray elevation angle and deviation from the trajectory can be leveraged for determining whether the shock wave originated from a cylindrical line source (90° ± 25°) or a spherical source (i.e., fragmentation episode). Atmospheric dynamics, including gravity waves, can affect infrasound propagation even at short ranges (Averbuch et al., 2022a; Averbuch et al., 2022b; Silber and Bowman, 2023), and therefore, it is sometimes necessary to include small-scale perturbations to atmospheric profiles (see Silber and Brown (2014) for more details).

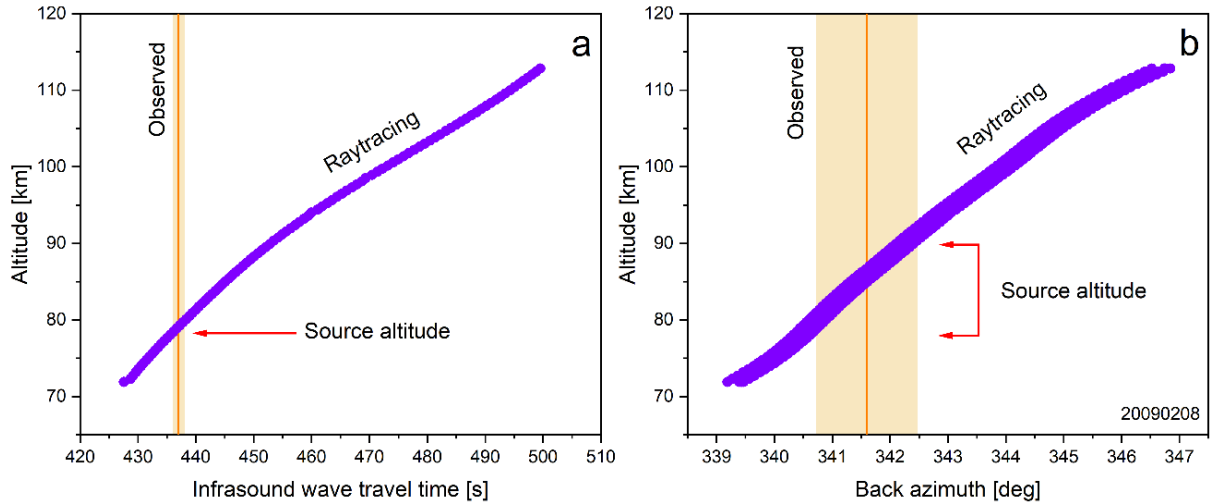

**Figure 8.** Example of raytracing results for an event with a well-constrained trajectory. The analysis of altitude versus infrasound wave travel time (left panel) and back azimuth (right panel) against observed values can be used to estimate the shock source height.

In far-field, it is generally assumed that the bolide trajectory can be approximated as a point-source (Edwards et al., 2006; Ens et al., 2012; ReVelle, 1976). While this might be the case, some caution must be exercised here. A significant consideration should be given to the bolide entry angle and the length of the propagation path. Shallow entry angles, especially if an event is very energetic, might translate into a source which can no longer be assumed to be strictly a point-source regardless of the distance. For example, Pilger et al. (2015) have found that infrasound signals generated by the Chelyabinsk event on the IMS stations showed evidence of acoustic emission directionality. This is not surprising given that the superbolide entered at an angle of only 16° (Peña-Asensio et al., 2022), thereby providing conditions such that the shock from the cylindrical line source will be produced over a long path length. As will be discussed in Section 5, the entry angle might be a possible cause of deviations in the observed back azimuths. This point has not been explored previously. On the other hand, steep entries can be approximated as a point source in terms of a geographical location. Here, however, the dominant effect will stem from the shock altitude and the mode of shock production. In particular, infrasound propagation waveguides might not even exist for sources at some altitudes but will at others. Therefore, propagation modeling can, in theory, provide constraints on shock source altitudes (Silber, 2024; Silber and Brown, 2014).

## 5. Case Study Analyses and Discussion

### 5.1 Near-field

The regional infrasound event (Ontario, Canada) had well-constrained ground truth, which was leveraged in searching for infrasound signals. Figure 9a shows the map with the full trajectory and the infrasound station. As described in Section 2, at short ranges, the geometry and length of the bolide trajectory relative to infrasound stations cannot be neglected. This event demonstrates such a scenario. The predicted back azimuth was 119° to 246° degrees, which is a large span to consider when searching for signals. Thus, the predicted infrasound travel times will vary accordingly. Infrasound from the regional event was detected on a local infrasound station just over 5 minutes after the onset of the luminous trail. The stacked and filtered (second order Butterworth filter with corner frequencies 0.3 – 8 Hz) waveform showing the signals is plotted in Figure 9b. The signal showed an N-wave signature, which is typical for near-field supersonic and hypersonic shock sources. The back azimuth indicated that the most likely source of the shock was the point indicated with the orange circle in Figure 9a.

A high value of trace velocity (0.8 km/s) was indicative of an airwave generated at an altitude and having a direct path to the station. Optical observations including the analysis of the light curve can provide evidence related to the physical conditions occurring during the meteoroid's flight (see Silber (2024) for more details). In particular, fragmentation episodes can be identified by examining points of sudden increase in brightness, thereby associating the type of shock to a cylindrical line source or a fragmentation episode. Aided by propagation modeling (see Section 4.3), the back azimuth and trace velocity were corroborated, and the shock source altitude was estimated at 88 km. This point is also shown in Figure 9c. From careful examination of the light curve, it also follows that the shock type was likely a cylindrical line source because there are no significant peaks in brightness (i.e., no fragmentation). The deviation angle of ~90° from the trajectory obtained from raytracing is also a diagnostic of a shock originating from a cylindrical line source. This event is a perfect example of observational and theoretical data fusion, and how well-constrained ground truth information can be leveraged for deriving source parameters.

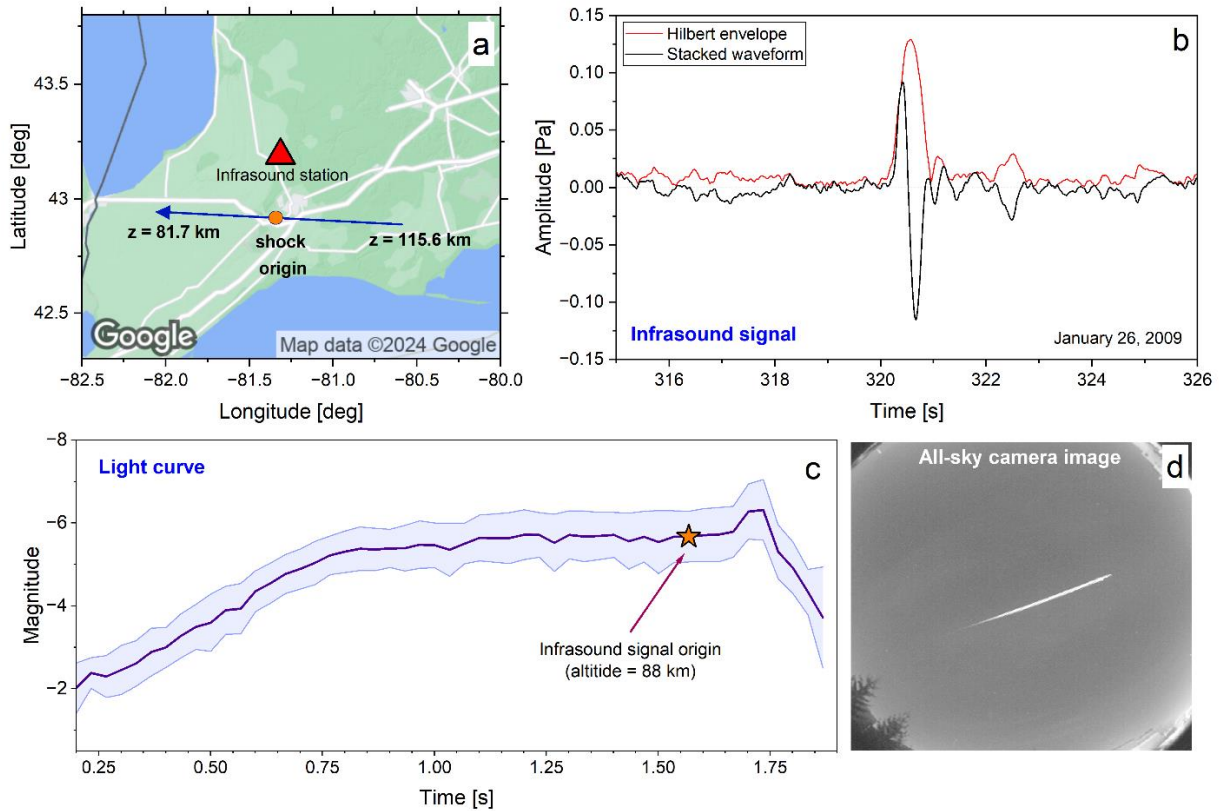

**Figure 9.** (a) Contextual map showing the event trajectory and infrasound station. The shock origin determined through propagation modeling and signal analysis is denoted with the orange circle. (b) Filtered 'stacked' waveform with the infrasound signal and the Hilbert envelope. The filter is 2nd order Butterworth, 0.3 – 8 Hz. The signal onset is at the ~320 second mark. The signal exhibits an N-wave signature. (c) Meteor light curve. The shaded region represents error in magnitude. The point at which infrasound was generated is annotated with the star. (d) The luminous trail of the meteor imaged with an all-sky camera (Image credit: The Meteor Physics Group, Western University). Map source: Google.

Infrasound generated by the Greenland bolide was analyzed in detail by Pilger et al. (2020). They leveraged complex infrasound signal features and the limited ground truth information to estimate source altitude and reconstruct the trajectory. Three IMS stations detected the signals, I18DK (65 km), I53US (2877 km), and I56US (3808 km). Figure 10a shows the location of I18DK and the location of bolide peak brightness (76.9°N, -69.0°E) as reported on the CNEOS website (orange star). The digitized light curve is shown in Figure 10b. The reported bolide velocity was 24 km/s, and an altitude of peak brightness was 43.3 km, which is slightly higher than the average for the entire CNEOS bolide population with known altitudes (Figure 11). The point of peak brightness was only 65 km from I18DK;

therefore, it is expected that the infrasound wave would reach it in approximately 5 minutes with a back azimuth of 176°. The bolide entered at a shallow angle of 11° which means the observed azimuth could deviate from the predicted value. In fact, Pilger et al. (2020) noted that the observed back azimuth at I53US was offset by 16°, although such deviations have been noted previously (e.g., Ott et al., 2019).

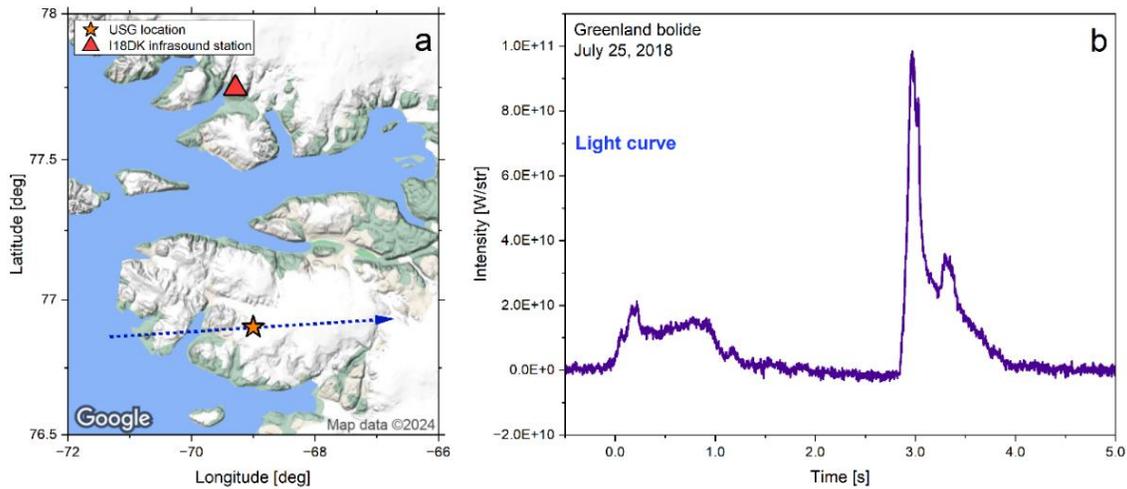

**Figure 10.** (a) Contextual map showing the location of the Greenland bolide and I18DK. The flight direction is shown with the arrow. The star indicates the point of peak brightness as reported on the CNEOS website. (b) The light curve generated by the USG sensors and digitized from a PDF file posted on the CNEOS website. Map source: Google.

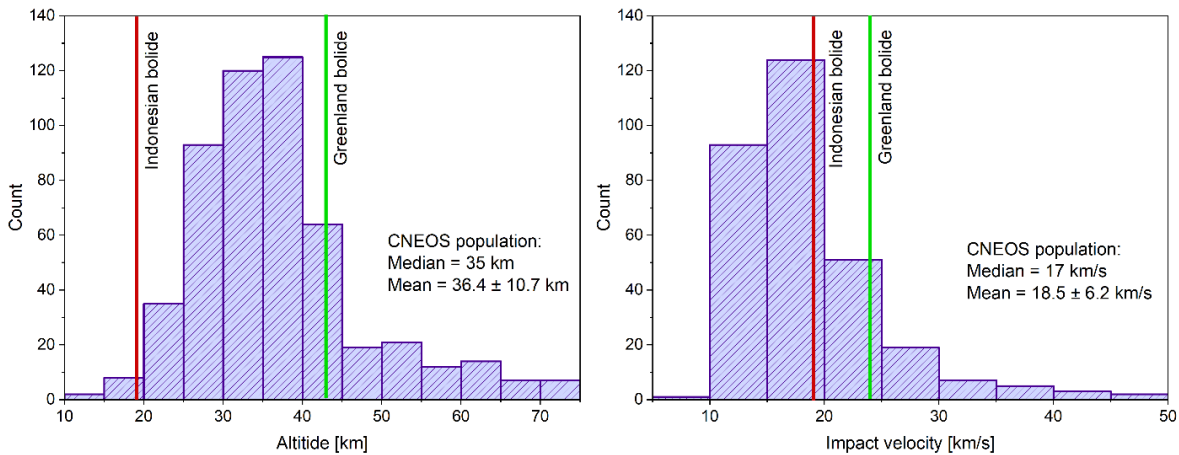

**Figure 11.** Population of all CNEOS bolides with known peak brightness altitudes and impact velocities. The Greenland and Indonesian bolides are shown with green and red lines, respectively.

The point of peak brightness corresponds to a gross fragmentation episode at the 3-second mark (Figure 10b), evidenced through a sudden increase in brightness. There was an additional, smaller fragmentation episode that followed. Indeed, locals reported hearing sonic booms, which had presumably originated from the bolide. Interestingly, there was an initial brightening (0-second mark), with much smaller peaks, followed by a slight decrease in brightness before gross fragmentation. This is likely indicative of strong ablation with small fragment shedding at higher altitudes. The combination of a shallow angle entry (more time spent in the atmosphere) and the fact that it exploded at an altitude of 43 km is suggestive of a weak impactor. Peña-Asensio et al. (2022) derived the orbital parameters and dynamic strength for most bolides in the CNEOS database. Based on the entry parameters, this particular object was likely of an asteroidal origin (Tisserand parameter = 3.07), but with low dynamic strength, which can explain why it fragmented at such an altitude. The diameter of the object was estimated to be ~2.5 meters.

The infrasound signal exhibited complex features and distinct packets which were interpreted to originate from a cylindrical line source, fragmentation, as well as air-to-ground coupling (see Pilger et al. (2020) for further details). The earliest signal arrived at 21:57:05 UTC with a trace velocity of 0.45 km/s. The initial arrivals were associated gross fragmentation, while later arrivals were associated with a line source as well as air-to-ground coupling. The total deviation in back azimuths across infrasound arrivals was 200°–240°, and 120°–170° for air-to-ground coupled waves. Considering that signal packets came from different apparent back azimuths, Pilger et al. (2020) were able to reconstruct a 3D trajectory. They estimated the entry angle and direction of propagation, respectively, at 3.1° and 77°. The CNEOS values are 11° for the entry angle and 81.3° for the direction of propagation. While it is not always possible to reconstruct the trajectory nor obtain signals from different types of shocks, this bolide event attests to the value of infrasound in obtaining source information that might not otherwise be available from other means of observation. Similar considerations apply to an artificial object re-entry, although some physical effects such as strong ablation might be absent. Therefore, natural objects can be used as analogues for studying artificial objects and vice versa (Silber et al., 2023). In terms of an artificial object re-entry, Genesis, Stardust, Hayabusa 1, Hayabusa 2 and OSIRIS-REx sample return capsules also exhibited N-wave signatures (Nishikawa et al., 2022; ReVelle et al., 2005; ReVelle and Edwards, 2006; Silber et al., 2023; Silber et al., 2024; Yamamoto et al., 2011). ReVelle et al. (2005) and ReVelle and Edwards (2006) leveraged the signal properties to derive the source energy for Genesis and Stardust, respectively. However, fragmentation, even if it is undesirable, can occur during a re-entry (e.g., space debris or a sample return capsule). An example of a re-entry with a fragmentation event is Hayabusa 1 (Yamamoto et al., 2011).

### 5.2 Far-field

The Indonesian bolide (8 October 2009) was detected at numerous stations of the IMS. The CNEOS database was not released until a few years after the event and only infrasound records were available at the time. Thus, ground truth information was limited to only witness reports. The initial analysis and signal measurements indicated an energetic source (Silber et al., 2011). As noted in the earlier sections, the size of the blast radius, and therefore the dominant signal period is related to energy deposition. However, the altitude effect can also play a role in the observed signal periods, as the relationship between $R_0$ and ambient pressure is inversely proportional (see Eqs. (1)-(2)). The competing process between signals emanating from high versus low altitude on one hand, and those produced by fragmentation episodes (including an airburst) on the other, can impede our ability to determine the cause for varying signal periods.

Silber et al. (2011) found signals at 17 IMS stations up to 17500 km from the reported source, with signal periods exhibiting large variations from a station to a station (5.8–25.2 s) (Figure 12). The closest station was I39PW (~2100 km) and the most distant was I08BO. Despite the event being very energetic, some signals were barely discernable either because they were weak or because the station was noisy, or a combination of both. Through a combination of numerical modeling and the observed signal analysis at stations up to 5000 km from the source, Silber et al. (2011) derived the source parameters and concluded that the longer signal periods were likely produced by a low altitude (~20 km) airburst, while shorter periods were linked to energy deposition by the hypersonic passage. The USG sensor data released later indicated that the peak brightness altitude was 19.1 km.

The subsequent re-analyses were performed by Gi and Brown (2017) and Ott et al. (2019), who also found large variations in the signal periods across stations. Gi and Brown (2017) found detections at 12 IMS stations at distances up to 12700 km from the source (the most distant station was I56US). Ott et al. (2019) found detections at 14 IMS stations at distances up to 14000 km (the most distant station was I17CL). They also noted that some signals were unreliable because of a very low SNR and excluded them from their energy calculations (I30JP, I46RU, I26DE, I18DK, and I56US). Neither study could find any detections at I08BO.

As mentioned earlier, this event was selected not only because it was energetic but to demonstrate possible challenges when it comes to infrasound signal analysis, especially if the signal SNR is low. It is interesting to note that the signal period measurements from the three published studies exhibited significant variations (Figure 12).

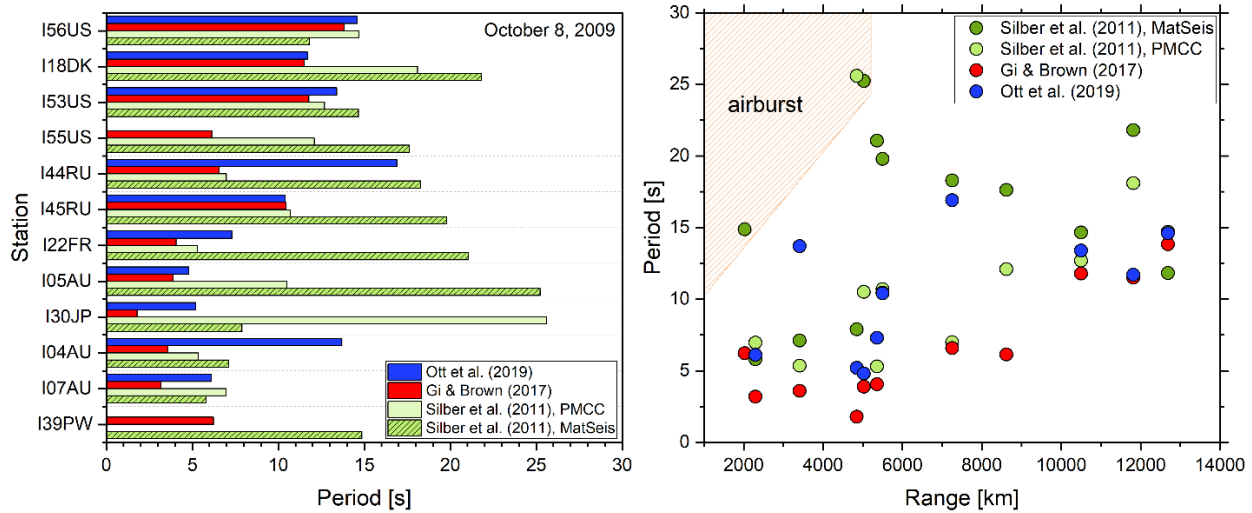

**Figure 12.** Signal periods reported in the literature (Gi and Brown, 2017; Ott et al., 2019; Silber et al., 2011). Left: Signal periods recorded at various IMS stations. Right: Signal periods as a function of range. The shaded region represents the signal periods that Silber et al. (2011) associated to the airburst.

Some variations could be due to station noise as several have been noted to have a poor SNR. Moreover, signal periods will be longer as a function of distance due to the attenuation of higher frequencies with range; this pattern can be seen in Figure 12 (right panel). Although the scatter in signal periods can occur as a result of station noise and as previously mentioned, propagation and distance effects, it is also a diagnostic of signals being generated at different parts of the trail and at different altitudes.

However, when comparing the three studies, i.e., signal period at any one given station, agreement in measurements was found only at a few stations. For example, at I04AU, the period measured by Silber et al. (2011), Gi and Brown (2017) and Ott et al. (2019) was 7.1 s, 3.6 s, and 13.7 s, respectively. Silber et al. (2011) applied two approaches; one involved measurement using the method described in Section 4, and another, using PMCC. PMCC-derived periods were generally smaller except in one case. The signal periods reported by Gi and Brown (2017) and Ott et al. (2019) also showed notable variations across different stations. The apparent differences in outcomes could be attributed to the variations in software packages used, parameter fine-tuning, sensitivity of the signal to parameters used, and poor SNR which can lead to a more subjective analysis. This bolide highlights the challenges in achieving consistent results, indicating that sometimes there may not be a single definitive solution, requiring further investigation. Future research should delve deeper into the causes of signal period variations and address analysis inconsistencies.

To ensure consistency in deriving the residuals (the difference between predicted and observed back azimuths) and comparing different studies, the predicted back azimuth is based on the location of peak brightness as listed in the CNEOS database. Excluding two stations (I56US and I39PW), the observed back azimuths were mostly consistent with that predicted. I56US, and especially I39PW exhibited large back azimuth residuals, 26°–29° and 34°–44°, respectively (Figure 13, left panel). It might be possible, although unlikely, that this was due to propagation effects. The three studies pointed out that the signal at I39PW was very weak, and it is plausible that beamforming was adversely affected by the lack of reasonably strong signals. Also, the station might have captured the signal when the bolide was at a much higher altitude, which could also explain the weak signals at I39PW but that still cannot account for such large back azimuth residuals. Similar to the signal periods (Figure 12), different studies obtained slightly different values of back azimuths. The most notable difference is seen in I55US, I18DK and I39PW (Figure 13, left panel).

One important question to answer is how much variability is there in predicted versus observed back azimuths. The fact that there are stations exhibiting large residuals indicate that the Indonesian bolide is unlikely to be an isolated case. Since the Chelyabinsk bolide, as the most energetic event recorded by the IMS network, was detected globally by a large number of stations, the residuals in the observed back azimuths should be compared and contrasted to those of the Indonesian bolide. Figure 13 (right panel) displays the back azimuth residuals from studies of Gi and Brown (2017) and Ott et al. (2019) for the two bolide events. The back azimuth residuals are notably greater in the Chelyabinsk bolide, readily approaching 20°. The median for the Indonesian bolide is 1.5°–2°, while the median for the Chelyabinsk bolide is 7°–9°. The back azimuth residual of 74° at I34MN is an outlier, and although shown in the plot, it was excluded from the median calculation. It is important to note that the Chelyabinsk superbolide had an exceptionally high yield (440 kt), depositing large amounts of energy along its long path and resulting in persistently strong signals. Other than a significant difference in energy deposition, a notable disparity between the two events is the entry angle. The Chelyabinsk bolide entered at a shallow angle (16°), while the Indonesian bolide entered at a fairly steep angle (67°). Bolides entering at a shallow angle will consequently exhibit some acoustic energy directionality as noted by Pilger et al. (2015). They found evidence of an acoustic radiation directionality associated with the Chelyabinsk bolide, favoring a cylindrical line source geometry. Although bolides can be assumed to be point-sources at very large distances (ReVelle, 1976; Silber, 2024), this might not be strictly so for very energetic events entering at a shallow angle.

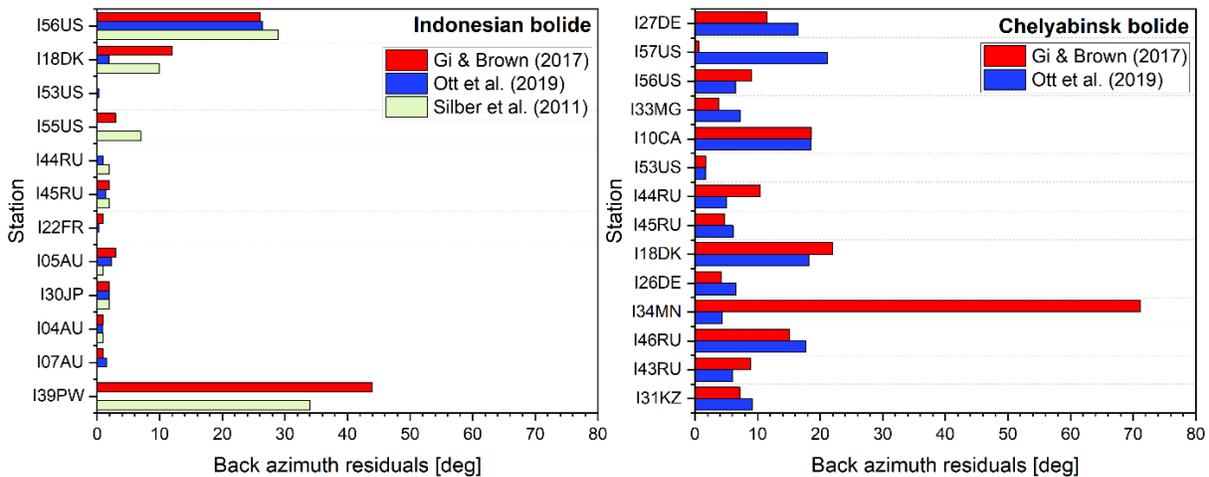

**Figure 13:** Observed back azimuth residuals for the Indonesian bolide (left) and the Chelyabinsk bolide (right). The median for the Indonesian bolide is 1.5° - 2°, and for the Chelyabinsk bolide 7° - 9°.

Future studies should investigate and test for all possible causes of notable (>10°) deviations in the observed back azimuths using a large sample of energetic events (low energy events are less likely to generate signals strong enough to reach a sufficiently large number of stations), while also accounting for propagation effects (Drob et al., 2010). Consideration should also be given to the directionality of acoustic emissions and the mode of shock production (Pilger et al., 2019; Silber, 2024). Infrasound generated by bolides entering at a steep entry angle might more readily propagate through atmospheric waveguides irrespective of the mode of shock production because of favorable geometry (Ronac Giannone et al., 2023).

## 6. Conclusions and Future Direction

Infrasound sensing is crucial for the detection and analysis of bolides, offering passive and cost-effective global monitoring capabilities. This review paper presents a comprehensive overview of the characteristics that differentiate bolides from other infrasound sources and aims to contribute to a more nuanced understanding of infrasound signal characteristics from bolides. The global monitoring capabilities of infrasound, as part of the IMS, underline its importance in complementing other sensing modalities as well as providing independent observations. The challenges presented by the intrinsic nature of bolides, including their passage through different atmospheric layers at hypervelocity and the often poorly constrained initial parameters, require a systematic and holistic approach to infrasound processing. This topic is also relevant to planetary exploration and space missions, particularly in the context of atmosphere-bearing planetary bodies, such as Mars, Titan, and Venus.

Three representative case studies have been presented to illustrate the practical application of infrasound processing methodologies, and different considerations when detections are made in near- and far-field. These studies highlight the utility of infrasound in deriving source parameters and demonstrate the challenges in interpretation, including an apparent disparity in signal analysis across different studies (e.g., signal period measurements). While there are many possible avenues of investigation, in terms of bolide infrasound processing and interpretation, the pressing topics are: (1) variability in the observed back azimuths and (2) intrinsic and extrinsic variability in signal periods. These are critical for reliable event geolocation and yield estimates, respectively. Therefore, the recommendations for future work are as follows (note that this is not an exhaustive list):

(1) Rigorously and systematically examine as well as test for all potential causes of variability in the observed signal periods.

(2) From (1), determine if a correction can be applied to new events to refine signal period estimates and improve yield calculations using period-based energy relations, considering factors like Doppler shift, propagation effects, and station effects.

(3) Develop strategies to address inconsistencies in the signal measurement, analysis and interpretation.

(4) Investigate possible causes for large (>10°) deviations in the observed back azimuths, while also taking into consideration variability in propagation effects. Ideally, this should be done using a large sample of reasonably energetic events (low energy events are less likely to generate signals strong enough to reach a sufficiently large number of stations).

(5) Expand on to-date work and statistically explore infrasound records associated with well-documented and well-characterized events toward better understanding of bolide mass distribution and impact flux.

In general, future work should continue to build on the unified framework presented in this paper, exploring additional case studies and refining techniques to accommodate the evolving nature of infrasound sensing technology. Enhancing the understanding and processing of infrasound signals generated by bolides and artificial re-entry events has the potential to advance impact hazard assessment and planetary defense initiatives.

**Funding:** This work was supported by the Laboratory Directed Research and Development (LDRD) program at Sandia National Laboratories, a multimission laboratory managed and operated by National Technology and Engineering Solutions of Sandia, LLC., a wholly owned subsidiary of Honeywell International, Inc., for the U.S. Department of Energy's National Nuclear Security Administration under contract DE-NA0003525.

**Data Availability Statement:** All data used in this paper were previously published (see references throughout). The CTBT IMS infrasound data are restricted but can be requested directly from CTBTO through the virtual Data Exploitation Centre (vDEC). Data from a small subset of CTBT IMS infrasound stations are freely available through EarthScope Data Services: https://ds.iris.edu/ds/.

**Acknowledgments:** The author thanks Ben Fernando and an anonymous reviewer for their insightful comments that helped improve the manuscript. The author gratefully acknowledges Brent Haglund (Sandia's Creative Services) for his graphic design contributions to Figures 1 and 2. Sandia National Laboratories is a multi-mission laboratory managed and operated by National Technology and Engineering Solutions of Sandia, LLC (NTESS), a wholly owned subsidiary of Honeywell International Inc., for the U.S. Department of Energy's National Nuclear Security Administration (DOE/NNSA) under contract DE-NA0003525. This written work is authored by an employee of NTESS. The employee, not NTESS, owns the right, title, and interest in and to the written work and is responsible for its contents. Any subjective views or opinions that might be expressed in the written work do not necessarily represent the views of the U.S. Government. The publisher acknowledges that the U.S. Government retains a non-exclusive, paid-up, irrevocable, world-wide license to publish or reproduce the published form of this written work or allow others to do so, for U.S. Government purposes. The DOE will provide public access to results of federally sponsored research in accordance with the DOE Public Access Plan.

**Conflicts of Interest:** The author declares no conflicts of interest.